\documentclass[aps,prb,twocolumn,superscriptaddress,longbibliography]{revtex4-2}
\usepackage{graphicx,color}
\usepackage{amsthm}
\usepackage{amsfonts}
\usepackage{algorithmic}
\usepackage{enumerate}
\usepackage{latexsym}
\usepackage{amsmath}
\usepackage{amssymb}
\usepackage{graphicx}
\usepackage[colorlinks=true,citecolor=blue,linkcolor=blue, urlcolor=blue]{hyperref}

\emergencystretch\maxdimen
\begin{document}
	
\title{Pairing Symmetry Crossover from $d$-wave to $s_{\pm}$-wave in a Bilayer Nickelate Driven by Hund's Coupling and Crystal Field Splitting}
	
\author{Yicheng Xiong}
\affiliation{School of Physics and Astronomy, Beijing Normal University, Beijing 100875, China\\}
	
\author{Yanmei Cai}
\affiliation{School of Physics and Astronomy, Beijing Normal University, Beijing 100875, China\\}

\author{Tianxing Ma}
\email{txma@bnu.edu.cn}
\affiliation{School of Physics and Astronomy, Beijing Normal University, Beijing 100875, China\\}
\affiliation{Key Laboratory of Multiscale Spin Physics (Ministry of Education), Beijing Normal University, Beijing 100875, China\\}
	
\begin{abstract}
The pairing symmetry of the recently discovered bilayer nickelate superconductor $\text{La}_3\text{Ni}_2\text{O}_7$ is a subject of intense debate in condensed matter physics, with the two leading theoretical candidates being a sign-reversing $s_{\pm}$-wave and a $d$-wave state. To investigate its ground-state properties in the intermediate coupling regime which is critical for real materials, we construct a two-orbital bilayer Hubbard model and employ the constrained-path quantum Monte Carlo method for large-scale simulations. 
By systematically calculating ground-state pairing correlation functions across parameter spaces, we map its pairing symmetry phase diagram. 
We find that an increasing Hund's coupling selectively enhances the interlayer $s_{\pm}$-wave pairing while suppressing the intralayer $d$-wave pairing. Similarly, a larger crystal field splitting drives a transition from $d$-wave- to $s_{\pm}$-wave-dominant states. Further analysis reveals that the strength of the intralayer $d$-wave pairing is strongly correlated with the $(\pi, \pi)$ antiferromagnetic spin fluctuations, which are in turn effectively suppressed by a large crystal field splitting, thereby weakening the $d$-wave pairing channel. Additionally, the dominant pairing symmetry transition region roughly overlaps with the inversion of orbital occupancy response to Hubbard $U$, suggesting an intrinsic link between pairing competition and orbital physics. Our results indicate that, within the parameter regime relevant to the actual material, the $s_{\pm}$-wave is the most probable pairing symmetry.
\end{abstract}
\maketitle
	
\noindent
\underline{\it Introduction}
The recent discovery of a superconducting (SC) transition temperature ($T_c$) of up to 80 K in the bilayer nickelate $\text{La}_3\text{Ni}_2\text{O}_7$ under high pressure has attracted significant attention\cite{sunSignaturesSuperconductivity802023}. This discovery establishes $\text{La}_3\text{Ni}_2\text{O}_7$ as a new member of the high-$T_c$ nickelate family. The low-energy physics of $\text{La}_3\text{Ni}_2\text{O}_7$ exhibits pronounced orbital selectivity: in contrast to infinite-layer nickelates, the correlation effects are stronger in the $\text{Ni-}d_{3z^2-r^2}$ orbital than in the $d_{x^2-y^2}$ orbital. Experimentally, angle-resolved photoemission spectroscopy (ARPES) measurements\cite{yangOrbitaldependentElectronCorrelation2024} have indicated a smaller mass renormalization factor for the $d_{x^2-y^2}$ orbital. This conclusion of differing correlation strengths between the two orbitals is also supported by DFT+DMFT calculations\cite{caoFlatBandsPromoted2024, ouyangHundElectronicCorrelation2024a} and infrared optical conductivity experiments\cite{liuElectronicCorrelationsPartial2024a}. This characteristic underscores the crucial role of Hund's coupling and multiorbital physics.
	
Given the complex multiorbital and correlated nature of $\text{La}_3\text{Ni}_2\text{O}_7$, numerous theoretical efforts have been devoted to understanding its SC mechanism. However, the pairing symmetry remains a subject of debate, with ongoing controversy between theoretical predictions and experimental results. Theoretically, weak-coupling approaches, such as the Random Phase Approximation (RPA)\cite{botzelTheoryMagneticExcitations2024a, lechermannElectronicCorrelationsSuperconducting2023, liuWavePairingDestructive2023, xiaSensitiveDependencePairing2025a, zhangStructuralPhaseTransition2024b} and the Functional Renormalization Group (FRG)\cite{guEffectiveModelPairing2025a, yangPossibleWaveSuperconductivity2023}, as well as the bilayer $t$-$J$ model in the strong-coupling limit\cite{liaoElectronCorrelationsSuperconductivity2023b, luInterlayercouplingdrivenHightemperatureSuperconductivity2024, yangInterlayerValenceBonds2023}, predominantly point to a sign-changing $s$-wave pairing, i.e., $s_{\pm}$-wave\cite{guEffectiveModelPairing2025a, liuWavePairingDestructive2023,sakakibaraPossibleHighSuperconductivity2024, yangPossibleWaveSuperconductivity2023, zhangStructuralPhaseTransition2024b}, while, other theories have proposed that $d$-wave pairing\cite{heierCompetingPairingSymmetries2024, liuElectronicCorrelationsPartial2024a, xiaSensitiveDependencePairing2025a, xiTransitionWave22025a,liaoElectronCorrelationsSuperconductivity2023b, luInterlayercouplingdrivenHightemperatureSuperconductivity2024} could be dominant. On the experimental front, point-contact Andreev reflection (PCAR) on high-pressure bulk samples\cite{liuAndreevReflectionSuperconducting2025} has observed a prominent zero-bias conductance peak, a feature typically associated with a sign-changing gap such as in $d$-wave or $s_{\pm}$-wave pairing. In contrast, for thin film samples where superconductivity is realized via strain engineering, both scanning tunneling microscopy (STM)\cite{fanSuperconductingGapStructure2025} and ARPES\cite{shenNodelessSuperconductingGap2025} have clearly revealed a fully-gapped, nodeless SC gap on the Fermi surface. This result is more consistent with the picture of an anisotropic $s$-wave, particularly the $s_{\pm}$-wave. Therefore, although the nodeless $s_{\pm}$-wave is the prevailing candidate, direct evidence for nodes\cite{caoDirectObservationDwave2025} or a sign-changing gap\cite{fanSuperconductingGapStructure2025} also exists, 
and a definitive conclusion awaits more direct experimental verification and theoretical understanding.
	
Existing theoretical works, ranging from weak-coupling theories like RPA to the strong-coupling framework of the $t$-$J$ model, have provided valuable physical insights into the SC mechanism. Recent first-principles calculations\cite{yueCorrelatedElectronicStructures2025b} and experimental spectra\cite{geislerOpticalPropertiesElectronic2024, liuElectronicCorrelationsPartial2024a} indicate that the Coulomb interaction and bandwidth in this system are of the same order of magnitude, placing it in a delicate intermediate-coupling regime. In this regime, electron itinerancy and localization coexist\cite{chenElectronicMagneticExcitations2024, ouyangHundElectronicCorrelation2024a}, which suggests that a some numerical method capable of treating both aspects on an equal footing may offer a complementary and crucial perspective for unveiling the complete physical picture. In this work, we construct a two-orbital bilayer Hubbard model and employ the constrained-path quantum Monte Carlo (CPMC) method to study its ground-state properties. The advantage of the CPMC method is that it effectively mitigates the severe sign problem, allowing us to perform ground-state projection calculations on large-scale lattices; for more details, see the Appendix. 
We systematically scan key parameters: on-site Coulomb interaction $U$ from weak-to-intermediate coupling, Hund’s coupling ratio $J_H/U$, crystal field splitting $\Delta E$, and interlayer hopping $t_{\perp}$. By calculating ground-state pairing correlation functions for $d$-wave and $s_{\pm}$-wave channels, we map the pairing symmetry phase diagram and identify critical trends: pairing symmetry in $\text{La}_3\text{Ni}_2\text{O}_7$ is highly sensitive to $J_H$ and $\Delta E$—increasing either enhances interlayer $s_{\pm}$-wave pairing while suppressing intralayer $d$-wave pairing. Only under weak $J_H$ and strong electron correlation does dominance shift to $d$-wave; in material-relevant parameter regimes, $s_{\pm}$-wave is likely dominant, consistent with experimental hints of sign-changing pairing. Notably, larger $\Delta E$ weakens the $d$-wave channel by suppressing AFM fluctuations, enabling $d_{3z^2-r^2}$ orbital-dominated $s_{\pm}$-wave to prevail. Additionally, the pairing symmetry transition region overlaps with the inversion of orbital occupancy response to $U$, pointing to an intrinsic link between pairing competition and orbital physics. Our findings resolve conflicting theoretical predictions and provide quantitative guidance for tuning the SC state via external pressure or strain.

\noindent
\underline{\it Model and methods}	
To describe the low-energy physics of bilayer nickelates, we study a two-orbital Hamiltonian on a bilayer square lattice\cite{quBilayerModelMagnetically2024,luoBilayerTwoorbitalModel2023c,maCompetitionAntiferromagnetismSuperconductivity2023a, maDopingdrivenAntiferromagneticInsulatorsuperconductor2022a, maParametersDependentSuperconducting2025, moIntertwinedElectronPairing2025a, xiTransitionWave22025a} which consists of four parts:
\begin{equation}
H = H_{k\parallel} + H_{k\perp} + H_U + H_V.
\end{equation}
The specific form of each term is given as follows:
\begin{align}
H_{k\parallel}=&-\sum_{\langle i,j\rangle,\ell,\sigma} \Big( t_{1}^{x} c^{\dagger}_{i\ell x\sigma}c_{j\ell x\sigma} + t_{1}^{z} c^{\dagger}_{i\ell z\sigma}c_{j\ell z\sigma} + \text{h.c.}\Big) \nonumber \\
&-t_{\mathrm{hyb}}\sum_{i,\ell,\sigma} \Big(c^{\dagger}_{i\ell x\sigma}c_{i+\hat x,\ell, z,\sigma} - c^{\dagger}_{i\ell x\sigma}c_{i+\hat y,\ell, z,\sigma} + \text{h.c.}\Big) \nonumber \\
H_{k\perp}=&-t_{\perp}^{z}\sum_{i,\sigma} \Big(c^{\dagger}_{i,1, z,\sigma}c_{i,2, z,\sigma}+\text{h.c.}\Big)  \nonumber \\
H_U=&U\,\sum_{i,\ell,\alpha} n_{i\ell\alpha\uparrow}\,n_{i\ell\alpha\downarrow}+\sum_{i,\ell,\sigma,\sigma'}\bigl(U'-J_H\delta_{\sigma\sigma'}\bigr)\,n_{i\ell x\sigma}\,n_{i\ell z\sigma'}\nonumber \\
H_V=&-\mu\sum_{i,\ell,\alpha,\sigma} n_{i\ell\alpha\sigma} + \frac{\Delta E}{2}\sum_{i,\ell,\sigma} \Big(n_{i\ell x\sigma}-n_{i\ell z\sigma}\Big)  
\end{align}

\begin{figure}[tbp]
\includegraphics[scale=0.5]{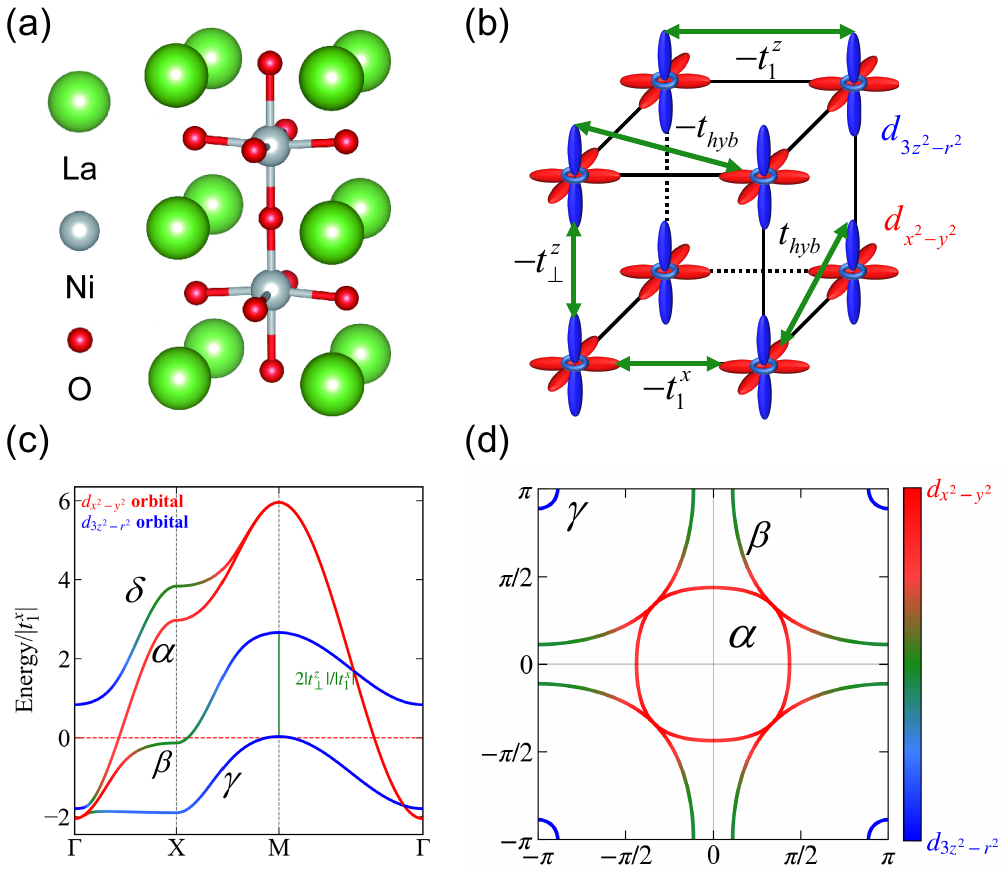}
\caption{(a) Crystal structure of $\text{La}_3\text{Ni}_2\text{O}_7$ under high pressure. (b) Schematic of the hopping processes in the model. (c) Band structure and (d)Fermi surface calculated using the tight-binding parameters from first-principles calculations for the material at 29.5 GPa (see Table~\ref{tab:model_parameters_combined}). Near the Fermi level, there are three bands $\alpha, \beta, \gamma$ and one unoccupied $\delta$ band. The splitting between the bonding and antibonding bands, arising from the interlayer coupling, is approximately $2t_\perp/t$ . The resulting Fermi surface consists of one electron-like pocket $\alpha$ at the Brillouin zone center ($\Gamma$ point) and two hole-like pockets $\beta, \gamma$ at the Brillouin zone corner ($M$ point). Here, red represents the $d_{x^2-y^2}$ orbital character and blue represents the $d_{3z^2-r^2}$ orbital character.}
\label{cpmcham}
\end{figure}
As that shown in Fig.~\ref{cpmcham}(a) and (b), the kinetic term $H_k = H_{k\parallel} + H_{k\perp}$ describes the electron hopping, which includes intra-layer nearest-neighbor intra-orbital hopping with amplitudes $t_1^x$ and $t_1^z$, inter-orbital hybridization $t_{\mathrm{hyb}}$, and inter-layer hopping $t_\perp^z$ that is restricted to the $d_{3z^2-r^2}$ orbital. We neglect next-nearest-neighbor hopping, which has been shown in many studies to not alter the essential physics\cite{luInterlayercouplingdrivenHightemperatureSuperconductivity2024, quBilayerModelMagnetically2024, yangInterlayerValenceBonds2023, zhangStructuralPhaseTransition2024b}.The on-site interaction term $H_U$ adopts a simplified Kanamori form, retaining only the density-density terms. The spin-flip and pair-hopping terms are neglected, as they have been shown to only slightly affect the phase diagram of a three-orbital Hubbard model\cite{liuOrbitalselectiveMottPhases2016}. Here, $U$ and $U'$ are the intra-orbital and inter-orbital Coulomb repulsions, respectively, and $J_H$ is the Hund's coupling. These parameters satisfy the relation $U'=U-2J_H$.  The potential energy term $H_V$ consists of the chemical potential $\mu$ and the crystal field splitting $\Delta E=\epsilon_x-\epsilon_z$ between the two orbitals.  In this work, we use $t \equiv t_1^x$ as the unit of energy and set $t=1$. For simplicity, we also denote $t_\perp \equiv t_\perp^z$.
	
Various studies have proposed that the following parameters have a significant influence on the SC properties and transition temperature of bilayer nickelates\cite{jiaSuperconductivityImbalancedBilayer2025, liaoElectronCorrelationsSuperconductivity2023b, luInterplayTwoOrbitals2024b, luoBilayerTwoorbitalModel2023c, moIntertwinedElectronPairing2025a, ouyangHundElectronicCorrelation2024a, shaoBandStructurePairing2025b, wangNormalSuperconductingProperties2024a, zhangElectronicStructureMagnetic2024, zhangGeneralTrendsElectronic2025, zhengWaveSuperconductivityBilayer2025,xiaSensitiveDependencePairing2025a}. To systematically investigate the key factors influencing SC pairing in this system, we perform a wide-range scan of the model parameters. The specific parameter ranges are listed in Table~\ref{tab:model_parameters_combined}. The range of $U/t$ covers the weak to intermediate electron correlation regime. The range of $J_H/U$ corresponds to typical values for the material and ensures the physical constraint of a positive inter-orbital interaction, $U'>0$. Concurrently, to simulate the physical scenarios of high-pressure bulk materials and ambient-pressure strained thin films, we tune $\Delta E/t$ and $t_\perp/t$. Previous studies have shown that the latter exhibits an approximately 30\% reduction in $t_\perp/t$ and a 40\% enhancement in $\Delta E/t$ compared to the former\cite{ushioTheoreticalStudyAmbient2025a, yueCorrelatedElectronicStructures2025b}. Our parameter ranges cover both types of materials.
	\begin{table}[htbp]
		\centering
		\begingroup
		\setlength{\tabcolsep}{8pt} 
		\begin{tabular}{cccc}
			\hline 
			\hline 
			$t_1^z/t$ & $t_{\mathrm{hyb}}/t$ & $t_\perp/t$ & $\Delta E/t$ \\
			\hline 
			$0.228$ & $-0.495$ & $1.315$ & $0.76$ \\
			\hline 
			$\boldsymbol{U/t}$ & $\boldsymbol{J_H/U}$ & $\boldsymbol{\Delta E/t}$ & $\boldsymbol{t_\perp/t}$ \\
			\hline 
			$1\sim6$ & $0.05\sim0.3$ & $0.45\sim0.9$ & $0.8\sim1.45$ \\
			\hline
			\hline  
		\end{tabular}
		\endgroup
		\caption{The parameters in regular font are the tight-binding parameters from first-principles calculations for the material under 29.5 GPa of pressure\cite{luoBilayerTwoorbitalModel2023c}. The parameters in bold indicate the ranges scanned in this work.}
		\label{tab:model_parameters_combined}
	\end{table}

To quantitatively evaluate the SC pairing tendency of the system, we calculate the pairing correlation function $C_\Gamma(\mathbf{R})$ for different symmetries $\Gamma$, defined as:
	$$C_\Gamma(\mathbf{R}) = \frac{1}{N_s} \sum_{i, \ell} \langle \Delta^\dagger_\Gamma(i+\mathbf{R}, \ell) \Delta_\Gamma(i, \ell) \rangle$$
	where $\Delta^\dagger_\Gamma$ is the order parameter for the corresponding symmetry. In this work, we primarily focus on two pairing channels: $d$-wave and $s_{\pm}$-wave. The order parameter for the $s_\pm$ state is positive on the $\beta$ sheet and negative on the $\alpha$ and $\gamma$ pockets\cite{botzelTheoryPotentialImpurity2025, qiuPairingSymmetrySuperconductivity2025, yangPossibleWaveSuperconductivity2023, zhangStructuralPhaseTransition2024b, zhanImpactNonlocalCoulomb2025}. This sign reversal is attributed to the difference in the bonding and antibonding character of the Fermi surface states on the $\beta$ and $\gamma$ pockets. In contrast, the $d$-wave order parameter has the same sign on all Fermi surface sheets. The specific forms of the order parameters are:
	\begin{align}
		\Delta^\dagger_{d}(i, \ell)& = \sum_{\delta} f(\delta) \left( c_{i\ell x\uparrow}^\dagger c_{i+\delta, \ell, x\downarrow}^\dagger - c_{i\ell x\downarrow}^\dagger c_{i+\delta, \ell, x\uparrow}^\dagger \right) \\
		\Delta^\dagger_{s \pm}(i)& = c_{i,1,z, \uparrow}^\dagger c_{i,2,z, \downarrow}^\dagger - c_{i,1,z, \downarrow}^\dagger c_{i,2,z, \uparrow}^\dagger
	\end{align}
	
	Here, the $d$-wave describes pairing between nearest-neighbor sites, with a form factor satisfying $f(\pm\hat{x})=+1$ and $f(\pm\hat{y})=-1$. To extract the true pairing strength, we subtract the uncorrelated single-particle contribution $C_\Gamma^{(0)}(\mathbf{R})$ from the total correlation. This term is calculated by decomposing the original four-fermion correlation function into a product of single-particle Green's functions\cite{huangAntiferromagneticallyOrderedMott2019h, maQuantumMonteCarlo2013e}. Finally, we obtain the effective pairing correlation $V_\Gamma$:
	\begin{equation}
		V_\Gamma = \sum_{\mathbf{R}}\big(C_\Gamma(\mathbf{R}) - C_\Gamma^{(0)}(\mathbf{R})\big)
	\end{equation}
	This quantity directly measures the pairing strength in the $\Gamma$ channel. A larger value of $V_\Gamma$ indicates a stronger tendency for pairing with that symmetry.
	
	To investigate the orbital selectivity and spin fluctuations in the bilayer nickelate system, we calculate a series of key physical quantities. First, to characterize the orbital-selective effects, we calculate the average electron occupancy $n_\alpha$ and the local double occupancy $D_\alpha$ for the $x$ and $z$ orbitals, respectively. Second, to probe the magnetic fluctuation characteristics of the system, we compute the static spin structure factor:
	\begin{equation}
		S(\mathbf{k}) = \frac{1}{N_s} \sum_{i,j}e^{i\mathbf{k}\cdot(\mathbf{R}_i-\mathbf{R}_j)} \langle S_i^z S_j^z \rangle
	\end{equation}
	The strength of AFM fluctuations in the system is measured by the peak intensity $S(\mathbf{M})$ at the wavevector $\mathbf{M}=(\pi, \pi)$. Additionally, we calculate the nearest-neighbor spin correlation function, $C_{nnspin} = \langle S_i^z S_{i+\delta}^z \rangle$. By combining these observables, we can perform a quantitative analysis of the electronic and magnetic ground-state properties of the system.

\noindent
\underline{\it Results}
	
\begin{figure}[tbp]
		\includegraphics[scale=0.38]{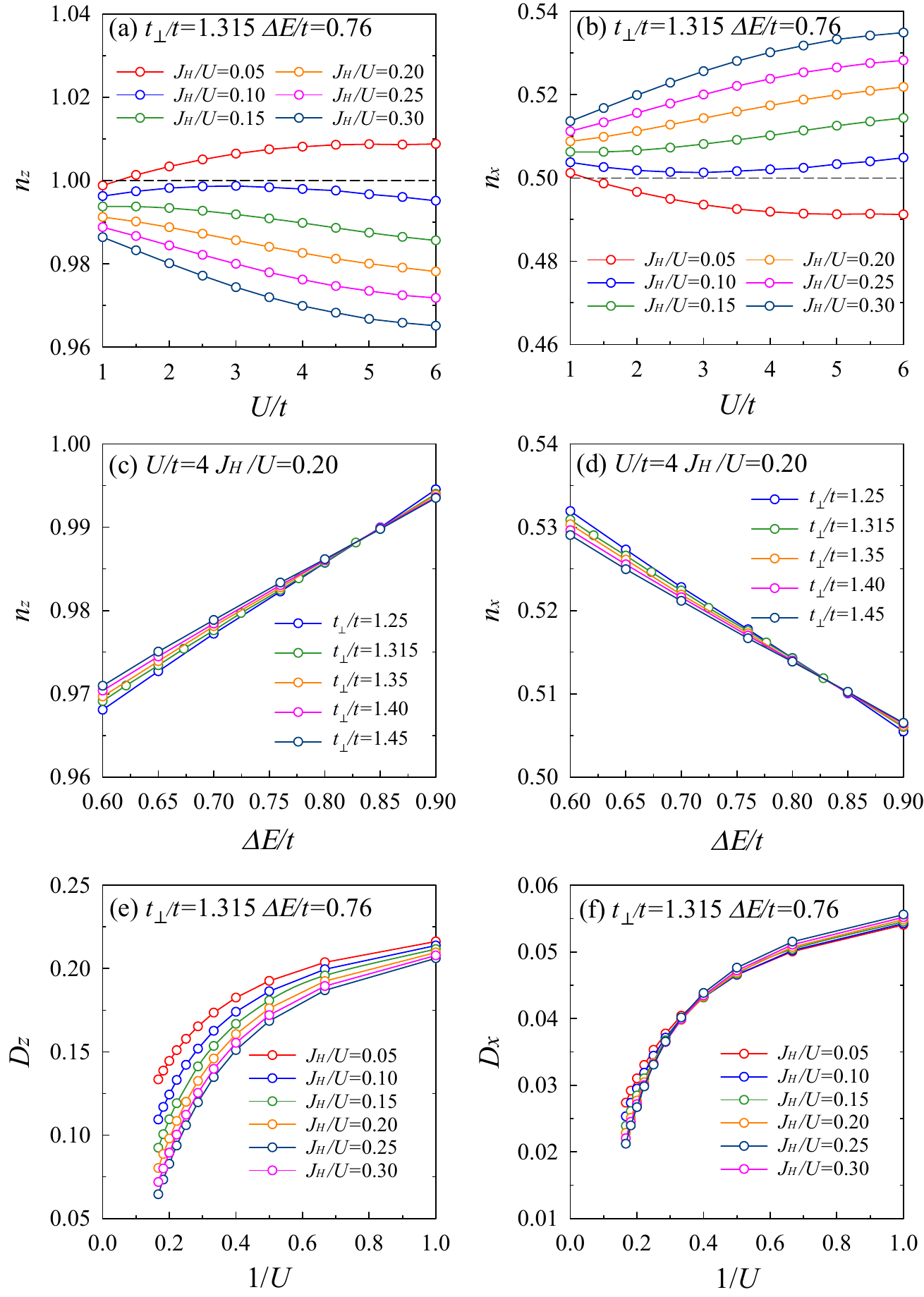}
		\caption{(a)(b) Evolution of electron occupancies $n_z$ and $n_x$ for the $z$ and $x$ orbitals as a function of the on-site Coulomb interaction $U/t$, at fixed interlayer coupling $t_{\perp}/t = 1.315$ and crystal field splitting $\Delta E/t = 0.76$. Curves of different colors correspond to different values of $J_H/U$. (c)(d) Orbital occupancies as a function of crystal field splitting $\Delta E/t$ at fixed $U/t=4$ and $J_H/U=0.20$. (e)(f) Intra-orbital double occupancies $D_z$ and $D_x$ for the two orbitals as a function of $1/U$. The total electron filling is fixed at $\langle n \rangle = 0.75$.}
		\label{orbitn}
	\end{figure}

	First, we investigate the effects of electronic correlations and lattice parameters on the $e_g$ orbital degrees of freedom. Figs. \ref{orbitn}(a) and \ref{orbitn}(b) illustrate the evolution of orbital occupancies with the Coulomb interaction $U$, the behavior of which depends on the Hund's coupling strength $J_H/U$. In the weak Hund's coupling regime ($J_H/U \lesssim 0.10$), increasing $U$ drives the transfer of electrons from the $x$ orbital to the lower-energy $z$ orbital, thereby enhancing orbital polarization. This effect can be understood as an effective enhancement of the crystal field splitting by the Coulomb interaction\cite{cartaExplicitDemonstrationEquivalence2025}. In the strong Hund's coupling regime, however, an increase in $U$ promotes electron flow to the $x$ orbital, thus weakening the orbital polarization. This reveals a competition between Hund's coupling, which favors uniform orbital occupation, and the crystal field splitting, which promotes orbital polarization. Notably, over a wide parameter space, our calculations show that the $x$ orbital remains nearly quarter-filled, while the $z$ orbital is close to half-filled. This robust occupation feature is consistent with theoretical predictions for Nd-doped systems\cite{chenElectronicStructuresSuperconductivity2025a} and observations from high-pressure XAS/XES experiments\cite{mijitStabilityLocalElectronic2025}. Figs. \ref{orbitn}(c) and \ref{orbitn}(d) further show that the orbital occupancies are predominantly determined by the crystal field splitting $\Delta E$, with $n_z$ increasing nearly linearly with $\Delta E$, while the influence of the interlayer coupling $t_{\perp}$ is minimal.
	
	The orbital-selective nature of electronic correlations is reflected in the behavior of the double occupancy $D$. As shown in Figs. \ref{orbitn}(e) and \ref{orbitn}(f), both $D_z$ and $D_x$ decrease monotonically with increasing $U$, indicating enhanced electron localization and a tendency towards a Mott insulating state. Interestingly, Hund's coupling significantly suppresses the double occupancy of the $z$ orbital, $D_z$. This is because Hund's rule promotes high-spin alignments of electrons in different orbitals, which effectively suppresses intra-orbital double occupancy in the nearly half-filled $z$ orbital. This effect is less pronounced for the $x$ orbital, reflecting an orbital-selective correlation effect.
	
\begin{figure}[tbp]
		\includegraphics[scale=0.4]{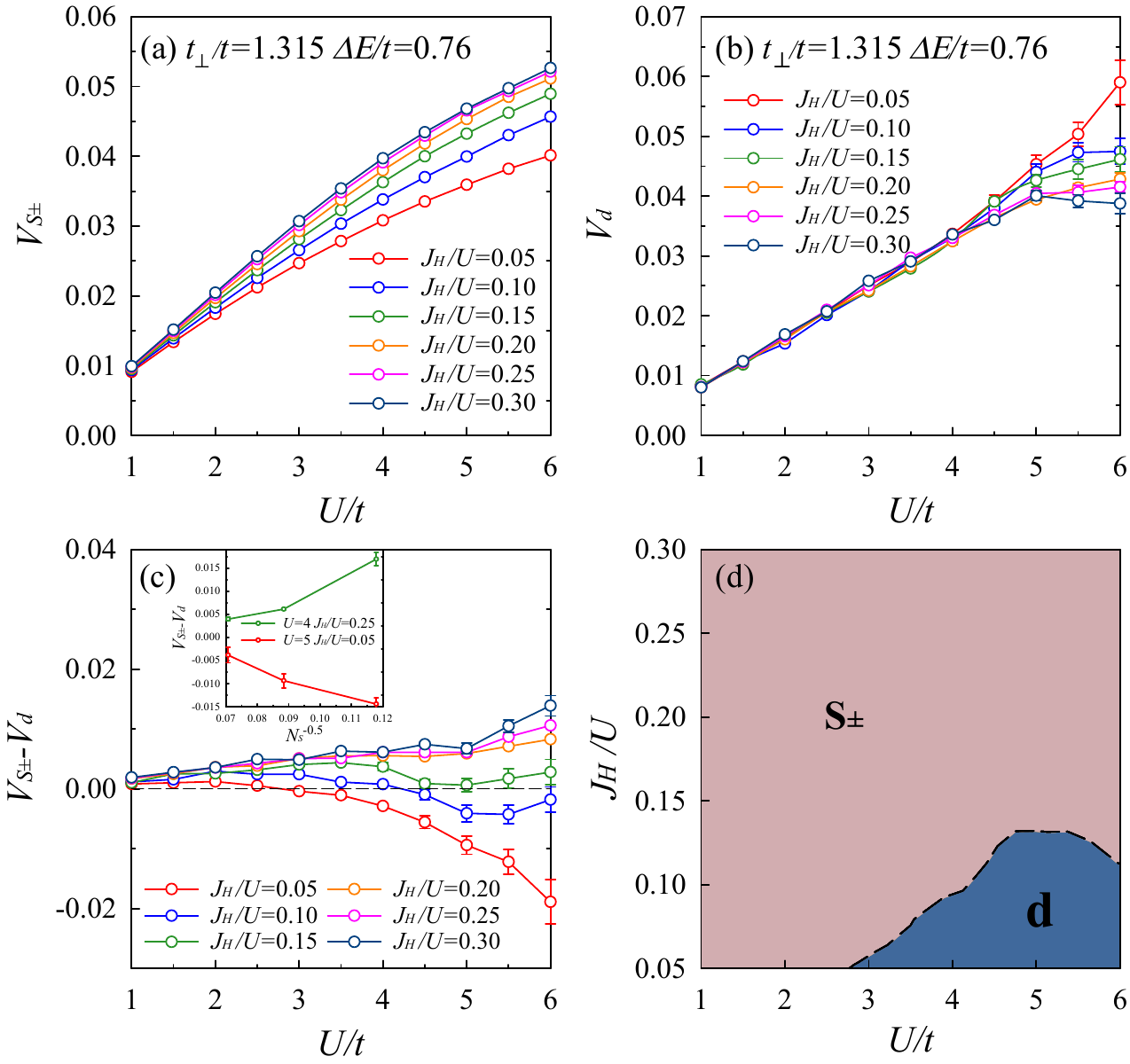}
		\caption{(a) $s_{\pm}$-wave pairing strength $V_{s_{\pm}}$ as a function of $U/t$ (parameters are $t_{\perp}/t = 1.315$, $\Delta E/t = 0.76$; different colors correspond to different $J_H/U$, same for below). (b) $d$-wave pairing strength $V_d$ as a function of $U/t$. (c) Evolution of the difference between the two pairing channel strengths, $V_{s_{\pm}}-V_d$, as a function of $U/t$. The inset shows the finite-size scaling analysis. (d) Phase diagram of the dominant pairing symmetry in the $U/t$–$J_H/U$ parameter plane: the pink region is $s_{\pm}$-wave dominant, the blue region is $d$-wave dominant, and the dashed line indicates the phase boundary ($V_{s_{\pm}} = V_d$).}
		\label{ujh}
	\end{figure}

	Next, we employ the CPMC method to calculate the effective pairing correlation $V$ and systematically analyze the influence of interaction parameters on the SCpairing symmetry. Fig. \ref{ujh}(a) shows that the $s_{\pm}$-wave pairing strength $V_{s_{\pm}}$ is monotonically enhanced with increasing $U$. At a fixed $U$, increasing $J_H$ further significantly boosts $V_{s_{\pm}}$, indicating that both Coulomb repulsion and Hund's coupling promote the $s_{\pm}$-wave pairing channel. In sharp contrast, Fig. \ref{ujh}(b) shows that while the $d$-wave pairing strength $V_d$ is enhanced by increasing $U$, it is suppressed by $J_H$. This suggests that $d$-wave pairing is favored in the weak Hund's coupling limit.
	
	Fig. \ref{ujh}(c) depicts the evolution of the difference in pairing strengths, $V_{s_{\pm}}-V_d$, which directly reflects the competition between the two pairing channels. For a small $J_H/U = 0.05$, the system undergoes a transition from being $s_{\pm}$-wave dominant to $d$-wave dominant as $U$ increases. As $J_H/U$ is increased, the entire curve of the difference shifts upwards. When $J_H/U \geq 0.15$, this difference remains positive over the entire range of $U$, indicating that a larger Hund's coupling effectively prevents the $d$-wave pairing from gaining dominance in the strong correlation regime, allowing the $s_{\pm}$-wave channel to remain dominant. To verify the robustness of this competitive trend in the thermodynamic limit, we performed a finite-size analysis for other system sizes ($N_{\text{states}}=144, L=6$ and $N_{\text{states}}=400, L=10$), as shown in the inset of Fig. \ref{ujh}(c). The results confirm that the different dominant pairing symmetries observed in different parameter regimes are not finite-size artifacts. Fig. \ref{ujh}(d) summarizes the pairing phase diagram in the $U-J_H$ plane. The results clearly indicate that Hund's coupling $J_H$ is a key parameter for tuning the pairing symmetry. Only in the region of weak $J_H$ and strong $U$ can the $d$-wave pairing, primarily driven by the $d_{x^2-y^2}$ orbital, potentially win. Interestingly, a comparison with Fig. \ref{orbitn}(a) shows that the region of the pairing symmetry transition is close to where the response of orbital occupancy to $U$ reverses, suggesting a possible intrinsic connection between the pairing competition and orbital physics.
	
\begin{figure}[tbp]
		\includegraphics[scale=0.4]{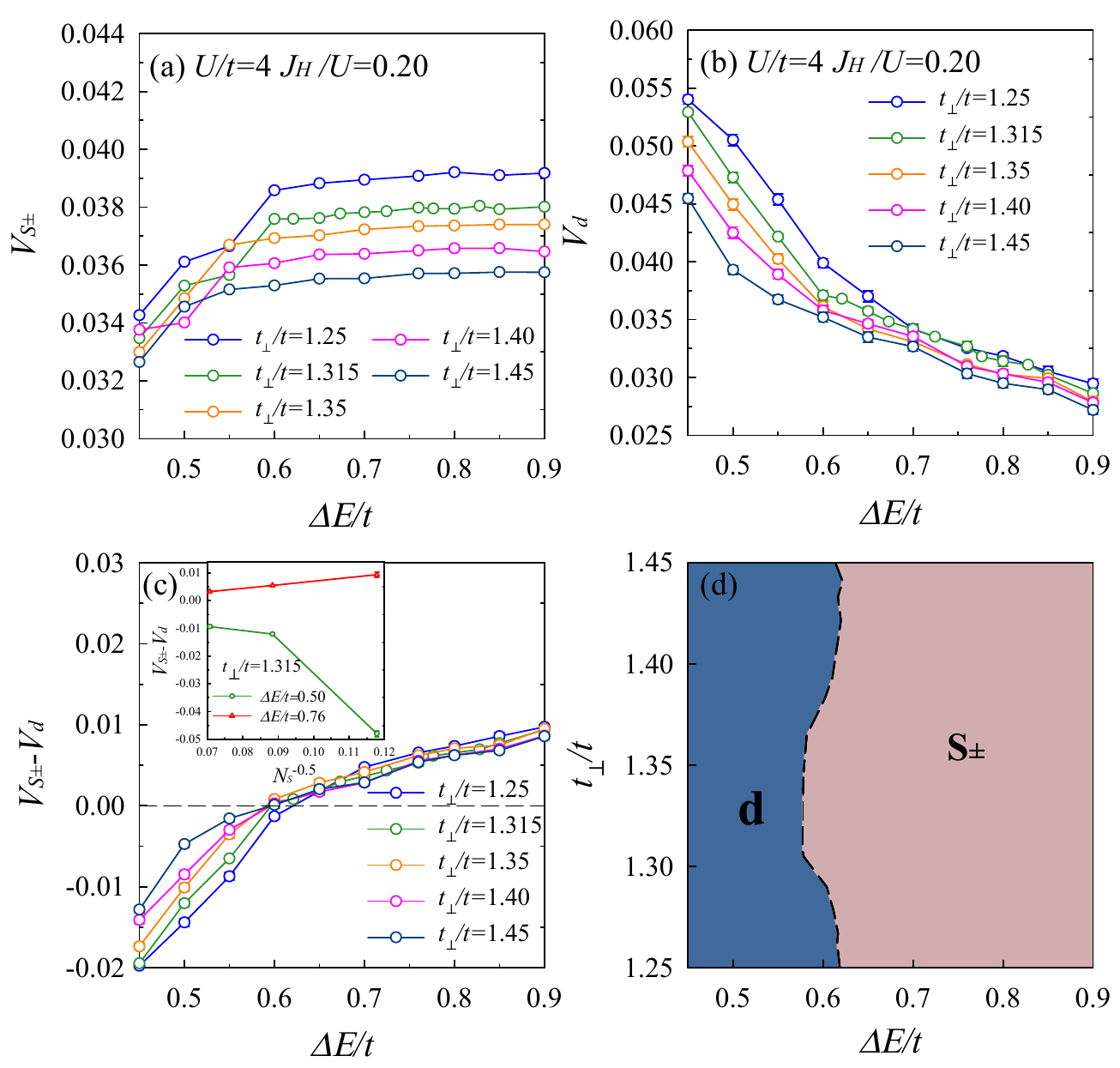}
		\caption{(a) $V_{s_{\pm}}$ as a function of $\Delta E/t$ (parameters are $U/t=4$, $J_H/U=0.20$; different curves correspond to different $t_{\perp}/t$, same for below). (b) $V_d$ as a function of $\Delta E/t$. (c) Evolution of the pairing potential difference $V_{s_{\pm}}-V_d$ as a function of $\Delta E/t$. (d) Phase diagram of the dominant pairing symmetry in the $\Delta E/t$–$t_{\perp}/t$ parameter plane: the pink region is $s_{\pm}$-wave dominant, the blue region is $d$-wave dominant, and the dashed line indicates the phase boundary.}
		\label{deltaetp}
	\end{figure}

	The lattice structural parameters, namely the crystal field splitting $\Delta E$ and the interlayer hopping $t_{\perp}$, can be tuned in the $\text{La}_3\text{Ni}_2\text{O}_7$ system by pressure or strain. Fig. \ref{deltaetp} demonstrates the impact of these two key parameters on the pairing competition. As shown in Figs. \ref{deltaetp}(a) and \ref{deltaetp}(b), increasing $\Delta E$ enhances the $s_{\pm}$-wave pairing while suppressing the $d$-wave pairing. The effect of the interlayer coupling $t_{\perp}$ is more complex. In the small $\Delta E$ region, increasing $t_{\perp}$ favors $d$-wave pairing, whereas in the large $\Delta E$ region, the influence of $t_{\perp}$ on the $d$-wave is minor. For the $s_{\pm}$-wave, decreasing $t_{\perp}$ generally favors pairing enhancement. This may be related to changes in the band structure; in this $t_{\perp}$ parameter region, a moderate decrease in $t_{\perp}$ raises the $\gamma$ band, which is mainly composed of the $d_{z^2}$ orbital, causing the $\gamma$ pocket to become larger and thereby enhancing the inter-Fermi-surface scattering that favors $s_{\pm}$-wave pairing. The detailed dependence on $t_{\perp}$ exhibits a dome-like shape, with specific data and analysis provided in the Appendix.
	
	The pairing strength difference in Fig. \ref{deltaetp}(c) indicates that as $\Delta E$ increases, the system undergoes a transition from $d$-wave to $s_{\pm}$-wave dominance, with the transition point located around $\Delta E/t \approx 0.6$. We have also performed a finite-size analysis in the inset, which shows that this competitive trend is robust in the thermodynamic limit. The structural parameter phase diagram in Fig. \ref{deltaetp}(d) summarizes this pattern: increasing the crystal field splitting is the primary driving force for the system to enter the $s_{\pm}$-wave pairing state, while the influence of the interlayer coupling $t_{\perp}$ on the phase boundary is relatively weak.
	
\begin{figure}[tbp]
		\includegraphics[scale=0.38]{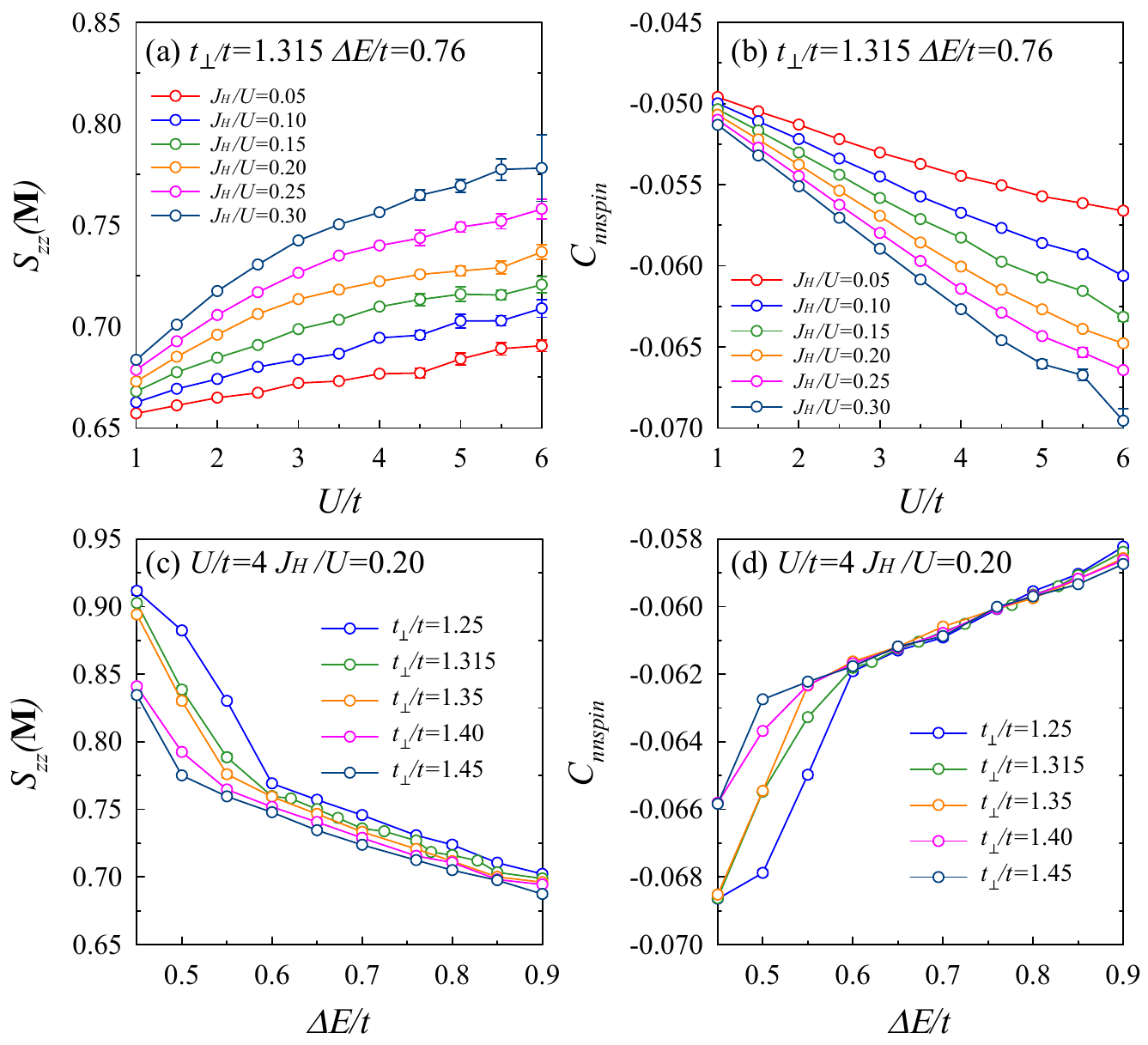}
		\caption{(a) $(\pi,\pi)$ static spin structure factor $S_{zz}(M)$ as a function of $U/t$ (parameters are $t_{\perp}/t=1.315$, $\Delta E/t = 0.76$; different colors correspond to different $J_H/U$, same for below). (b) Nearest-neighbor spin correlation $C_{nnspin}$ as a function of $U/t$. (c) $S_{zz}(M)$ as a function of $\Delta E/t$ (parameters are $U/t=4$, $J_H/U=0.20$; different colors correspond to different $t_{\perp}/t$, same for below). (d) $C_{nnspin}$ as a function of $\Delta E/t$.}
		\label{skm}
	\end{figure}

	Finally, we analyze the system's magnetic correlations. Figs. \ref{skm}(a) and \ref{skm}(b) show that increasing $U$ significantly enhances both the $(\pi,\pi)$ static spin structure factor $S_{zz}(M)$ and the nearest-neighbor AFM correlation $|C_{nnspin}|$. A comparison with Fig. \ref{ujh}(b) reveals that the evolution of the $d$-wave pairing strength is consistent with the trend of enhanced spin correlations with $U$, supporting the picture of spin fluctuations as the pairing mediator for $d$-wave pairing.
	
	The tuning of magnetism by structural parameters is shown in Figs. \ref{skm}(c) and \ref{skm}(d). Increasing the crystal field splitting $\Delta E$ significantly suppresses the AFM correlations. This trend is highly consistent with the behavior of the $d$-wave pairing, which weakens with increasing $\Delta E$ as shown in Fig. \ref{deltaetp}(b). In the small $\Delta E$ region, enhancing $t_{\perp}$ also suppresses the $(\pi, \pi)$ AFM correlation, while in the large $\Delta E$ region, $t_{\perp}$ has little effect on the AFM, which is also consistent with the response of $V_d$ in Fig. \ref{deltaetp}(b). Our results are consistent with recent theoretical predictions based on the random phase approximation (RPA)\cite{xiaSensitiveDependencePairing2025a}: increasing $\Delta E$ enhances the weight of the $d_{z^2}$ orbital component on the Fermi surface, alters the Fermi velocity, and suppresses the spin susceptibility at the Brillouin zone boundary (such as the M point), thereby leading to a transition from $d$-wave to $s_{\pm}$-wave pairing.
	
\noindent
\underline{\it Summary}
	
	In this paper, we have conducted a systematic numerical study of a two-orbital Hamiltonian model describing bilayer nickelate superconductors using the CPMC method. We focused on investigating the competition mechanism of SCpairing symmetries in this system, particularly the relationship between $d$-wave and $s_{\pm}$-wave pairings.
	
	By constructing pairing phase diagrams in the parameter space, our study finds that Hund's coupling $J_H$ and crystal field splitting $\Delta E$ are two key factors in tuning the pairing symmetry. Regarding electronic interactions, although the on-site Coulomb repulsion $U$ enhances both pairing channels, Hund's coupling $J_H$ exhibits a strong selectivity. It significantly promotes $s_{\pm}$-wave pairing while suppressing $d$-wave pairing. This characteristic allows $d$-wave pairing to become dominant only in the parameter region of weak Hund's coupling and strong correlation. Concerning the lattice structural parameters, increasing the crystal field splitting $\Delta E$ is the primary driving force for the transition from a $d$-wave dominant to an $s_{\pm}$-wave dominant state, whereas the influence of the interlayer coupling $t_{\perp}$ is relatively minor.
	
	To elucidate the physical origin of the aforementioned competition mechanism, we further analyzed the system's spin fluctuations. The results show that the strength of the intra-layer $d$-wave pairing is highly positively correlated with the strength of the $(\pi, \pi)$ AFM spin fluctuations. Both are enhanced with increasing $U$ and are significantly suppressed with increasing $\Delta E$. This result strongly supports the physical picture where spin fluctuations act as the pairing mediator for $d$-wave pairing. Therefore, a larger crystal field splitting $\Delta E$ weakens the $d$-wave channel by suppressing AFM fluctuations, which in turn allows the interlayer $s_{\pm}$-wave pairing, dominated by the $d_{3z^2-r^2}$ orbital, to take precedence. Furthermore, we also observed significant orbital-selective correlation effects, which are closely related to the complexity of the pairing behavior.

\noindent
\underline{\it Acknowledgments}
This work was supported by NSFC (12474218) and Beijing Natural Science Foundation (No. 1242022 and 1252022). The numerical simulations in this work were performed at the HSCC of Beijing Normal University.

	\appendix
	\section{CPMC METHODS}
	
	We employ CPMC method to investigate the magnetic and SCproperties of the system. This method projects a trial wave function $\Psi_T$ onto the ground state through a random walk in the space of Slater determinants via imaginary-time projection. This process can be expressed as:
	\begin{equation}
		\Psi_0 \propto \lim_{\Theta \to \infty} e^{-\Theta H}\,\Psi_T,
		\label{eq:projection}
	\end{equation}
	where $H$ is the Hamiltonian of the system and $\Psi_0$ is the ground-state wave function. When the projection is carried out with a finite imaginary-time step, the evolution at each step is governed by the short-time propagator:
	\begin{equation}
		|\Phi(\tau+\Delta\tau)\rangle = e^{-\Delta\tau H} |\Phi(\tau)\rangle.
		\label{eq:evolution}
	\end{equation}
	With a sufficiently large projection time $\Theta = M\Delta\tau$ (where $M$ is the number of projection steps), the excited-state components of $\Psi_T$ are progressively damped, allowing the wave function to converge to the ground state. In this process, the trial wave function is introduced as a reference for importance sampling. Through the constrained-path approximation, any random walk path that encounters a Slater determinant with a non-positive overlap with $\Psi_T$ is discarded, ensuring that only paths with positive overlap are accumulated. This is equivalent to imposing the condition that for any walker state $|\Phi\rangle$:
	\begin{equation}
		O_T(\Phi) = \langle \Psi_T | \Phi \rangle > 0;
		\label{eq:constraint}
	\end{equation}
	If a walker state results in $O_T(\Phi)\le 0$, its path is terminated, which effectively circumvents the fermion sign problem. This approximation renders the method variational: the calculated ground-state energy $E_{\rm CPMC}$ serves as an upper bound to the true ground-state energy $E_0$, with equality holding only if $\Psi_T$ is the exact ground-state wave function. Benchmark studies have shown that for models such as the Hubbard model, CPMC predictions for the ground-state energy and various physical observables are in excellent agreement with results from exact diagonalization and the density matrix renormalization group (DMRG)\cite{leblancSolutionsTwoDimensionalHubbard2015, qinAbsenceSuperconductivityPure2020}. Therefore, CPMC has been established as a reliable numerical tool for studying strongly correlated electron systems. For more technical details, see Refs.~\cite{zhangConstrainedPathMonte1997, zhangConstrainedPathQuantum1995}.
	
	In this work, we employ a trial wave function in the form of a single Slater determinant within the CPMC algorithm. By default, $\Psi_T$ is the ground state of the non-interacting single-particle Hamiltonian, which is a determinant constructed from the lowest-energy single-particle orbitals. Our simulations are performed with periodic boundary conditions. The number of random walkers is set to 1200, and the imaginary-time step is $\Delta\tau = 0.02$. After an initial thermalization period, we divide the Monte Carlo sampling into 40 blocks, with each block consisting of 320 projection steps, to ensure statistical independence between blocks. During the measurement phase, we average the observables over these blocks and estimate the statistical errors. We define the electron filling as $\langle n \rangle = N_{\mathrm{e}}/N_{\mathrm{orb}}$, which represents the average electron occupation per orbital (including spin). All simulations are performed at a fixed total electron filling of $\langle n \rangle = 0.75$. This corresponds to an average of 1.5 electrons in the $e_g$ orbitals per Ni site, consistent with the number of $e_g$ electrons for the Ni$^{2.5+}$ valence state in La$_3$Ni$_2$O$_7$.
	
	\section{Tuning Effect of $t_{\perp}$}
\begin{figure}[tbp]
		\includegraphics[scale=0.4]{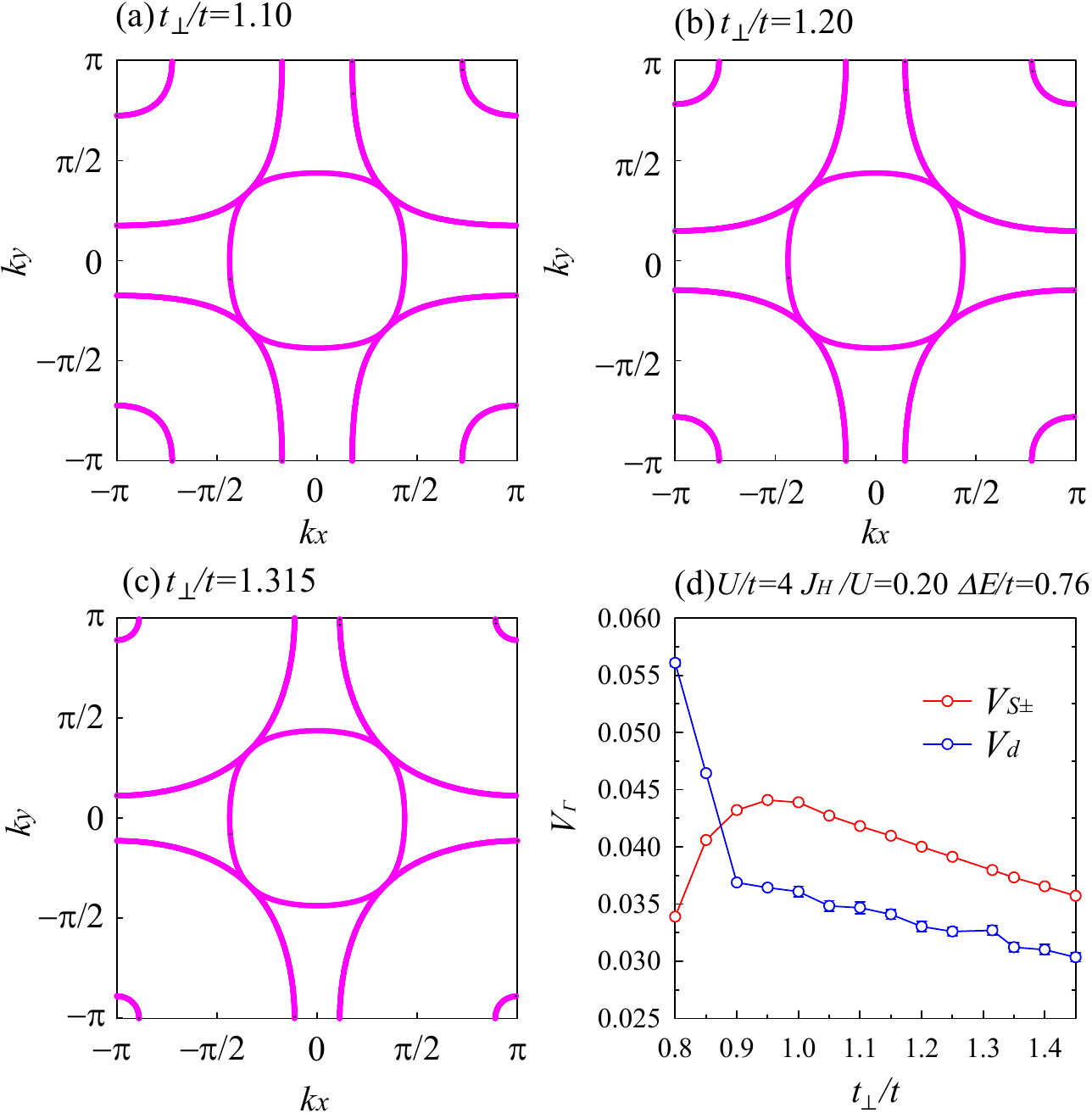}
		\caption{(a)-(c) Fermi surfaces in the non-interacting limit for $t_{\perp}/t = 1.1$, $1.2$, and $1.315$, respectively. (d) $V_{s_{\pm}}$ and $V_d$ as a function of $t_{\perp}/t$ at $U/t=4$, $J_H/U=0.20$, and $\Delta E/t=0.76$.}
		\label{tp}
	\end{figure}

	Within the parameter range studied, the interlayer hopping parameter $t_{\perp}$ consistently suppresses $d$-wave pairing. As shown in Fig.~\ref{tp}(d), the effective correlation function $V_d$, which represents the strength of $d$-wave pairing, decreases monotonically as $t_{\perp}/t$ increases. This indicates that enhanced interlayer coupling is detrimental to the formation of d-wave pairing. In contrast, the effect of $t_{\perp}/t$ on $s_{\pm}$-wave pairing is more complex, exhibiting a non-monotonic behavior: it is first enhanced and then suppressed, forming a "dome-like" dependence. In Fig.~\ref{tp}(d), $V_{s\pm}$, representing the strength of $s_{\pm}$-wave pairing, increases with $t_{\perp}/t$ in the small $t_{\perp}/t$ regime, reaches a peak, and is then suppressed in the large $t_{\perp}/t$ regime.
	
	Figs.~\ref{tp}(a)-(c) illustrate the evolution of the Fermi surface in the non-interacting limit as $t_{\perp}/t$ increases. In particular, as $t_{\perp}/t$ decreases from a large value,  suah as 1.315, the $\gamma$ pocket at the center of the Brillouin zone is observed to expand. This expansion may provide more phase space for inter-pocket scattering, thereby favoring the formation of $s_{\pm}$-wave pairing. In the small $t_{\perp}/t$ regime, although d-wave pairing is somewhat suppressed, it may still be the dominant pairing channel. However, as $t_{\perp}/t$ increases further, the $d$-wave pairing is rapidly weakened. Near a critical value of $t_{\perp}/t$, the system may undergo a transition in SCpairing symmetry from $d$-wave dominant to $s_{\pm}$-wave dominant. Some recent studies have also pointed to the possibility of such an interlayer-coupling-driven pairing symmetry transition in similar systems\cite{jiaSuperconductivityImbalancedBilayer2025}.
	
	\bibliography{Reference}
	
\end{document}